%% file: sig-alternate-sample.tex
\documentclass{sig-alternate-05-2015}

\begin{document}

% Copyright
\setcopyright{acmcopyright}
%\setcopyright{acmlicensed}
%\setcopyright{rightsretained}
%\setcopyright{usgov}
%\setcopyright{usgovmixed}
%\setcopyright{cagov}
%\setcopyright{cagovmixed}

\CopyrightYear{2016} 
\setcopyright{acmcopyright}
\conferenceinfo{SIGIR '16,}{July 17--21, 2016, Pisa, Italy.}
\isbn{978-1-4503-4069-4/16/07}\acmPrice{\$15.00}
\doi{http://dx.doi.org/10.1145/2911451.2914666}
\clubpenalty=10000 
\widowpenalty = 10000
%
% --- Author Metadata here ---
%\conferenceinfo{WOODSTOCK}{'97 El Paso, Texas USA}
%\conferenceinfo{SIGIR '16}{July, 17--21, 2016, Pisa, Italy}
%\CopyrightYear{2007} % Allows default copyright year (20XX) to be over-ridden - IF NEED BE.
\crdata{0-12345-67-8/90/01}  % Allows default copyright data (0-89791-88-6/97/05) to be over-ridden - IF NEED BE.
% --- End of Author Metadata ---

\title{A Cross-Platform Collection of Social Network Profiles}

%\title{Linked users across social network dataset collection \titlenote{(Produces the permission block, and
%copyright information). For use with
%SIG-ALTERNATE.CLS. Supported by ACM.}}
%\subtitle{[Extended Abstract]
%\titlenote{A full version of this paper is available as
%\textit{Author's Guide to Preparing ACM SIG Proceedings Using
%\LaTeX$2_\epsilon$\ and BibTeX} at
%\texttt{www.acm.org/eaddress.htm}}}
%
% You need the command \numberofauthors to handle the 'placement
% and alignment' of the authors beneath the title.
%
% For aesthetic reasons, we recommend 'three authors at a time'
% i.e. three 'name/affiliation blocks' be placed beneath the title.
%
% NOTE: You are NOT restricted in how many 'rows' of
% "name/affiliations" may appear. We just ask that you restrict
% the number of 'columns' to three.
%
% Because of the available 'opening page real-estate'
% we ask you to refrain from putting more than six authors
% (two rows with three columns) beneath the article title.
% More than six makes the first-page appear very cluttered indeed.
%
% Use the \alignauthor commands to handle the names
% and affiliations for an 'aesthetic maximum' of six authors.
% Add names, affiliations, addresses for
% the seventh etc. author(s) as the argument for the
% \additionalauthors command.
% These 'additional authors' will be output/set for you
% without further effort on your part as the last section in
% the body of your article BEFORE References or any Appendices.

\numberofauthors{2} %  in this sample file, there are a *total*
% of EIGHT authors. SIX appear on the 'first-page' (for formatting
% reasons) and the remaining two appear in the \additionalauthors section.
%
\author{
% You can go ahead and credit any number of authors here,
% e.g. one 'row of three' or two rows (consisting of one row of three
% and a second row of one, two or three).
%
% The command \alignauthor (no curly braces needed) should
% precede each author name, affiliation/snail-mail address and
% e-mail address. Additionally, tag each line of
% affiliation/address with \affaddr, and tag the
% e-mail address with \email.
%
% 1st. author
\alignauthor
Maria Han Veiga \\
       \affaddr{Inst. of Computational Science}\\
       \affaddr{University of Zurich, Switzerland}\\
       \email{hmaria@physik.uzh.ch}
% 2nd. author
\alignauthor
Carsten Eickhoff \\ 
       \affaddr{Dept. of Computer Science}\\
       \affaddr{ETH Zurich, Switzerland}\\
       \email{ecarsten@inf.ethz.ch}
% 3rd. author
%\alignauthor
%John Doe \\ %\titlenote{Dr.~Trovato insisted his name be first.}\\
%       \affaddr{Institute}\\
%       \affaddr{Street name}\\
%       \affaddr{City, Country}\\
%       \email{email@corporation.com}
%\and  % use '\and' if you need 'another row' of author names
% 4th. author
}
% There's nothing stopping you putting the seventh, eighth, etc.
% author on the opening page (as the 'third row') but we ask,
% for aesthetic reasons that you place these 'additional authors'
% in the \additional authors block, viz.
%\additionalauthors{Additional authors: John Smith (The Th{\o}rv{\"a}ld Group,
%email: {\texttt{jsmith@affiliation.org}}) and Julius P.~Kumquat
%(The Kumquat Consortium, email: {\texttt{jpkumquat@consortium.net}}).}
\date{10 February 2016}
% Just remember to make sure that the TOTAL number of authors
% is the number that will appear on the first page PLUS the
% number that will appear in the \additionalauthors section.

\maketitle
\begin{abstract}
%In this paper we present a unique data set which comprises of over 1000 users and their online social profiles in three distinct platforms (Twitter, Foursquare and Instagram). We describe the collection method and features of this data set. As the interest in the topic of user privacy is gaining traction, we make this data set available for public, so that it be used to validate methods in different problems, such as user identification or modelling.

The proliferation of Internet-enabled devices and services has led to a shifting balance between digital and analogue aspects of our everyday lives. In the face of this development there is a growing demand for the study of privacy hazards, the potential for unique user de-anonymization and information leakage between the various social media profiles many of us maintain. To enable the structured study of such adversarial effects, this paper presents a dedicated dataset of cross-platform social network personas (\textit{i.e.}, the same person has accounts on multiple platforms). The corpus comprises 850 users who generate predominantly English content. Each user object contains the online footprint of the same person in three distinct social networks: Twitter, Instagram and Foursquare. In total, it encompasses over 2.5M tweets, 340k check-ins and 42k Instagram posts. We describe the collection methodology, characteristics of the dataset, and how to obtain it. Finally, we discuss a common use case, cross-platform user identification.

%Our purpose is to make it accessible to the research community and to provide a benchmark dataset that researchers can use to study research questions which benefit from the availability of content generated by the same person in different social networks. The (LuaSN / L-users) dataset contains over 2.5 million Tweets, 350k check-ins and 42k Instagram posts. In this paper, we describe the collection methodology, characteristics of the dataset, how to obtain it and we show one of its use cases, in user identification.

\end{abstract}

%
% The code below should be generated by the tool at
% http://dl.acm.org/ccs.cfm
% Please copy and paste the code instead of the example below. 
%
\begin{CCSXML}
<ccs2012>
<concept>
<concept_id>10002951.10003317.10003359.10003360</concept_id>
<concept_desc>Information systems~Test collections</concept_desc>
<concept_significance>300</concept_significance>
</concept>
</ccs2012>
\end{CCSXML}

\ccsdesc[300]{Information systems~Test collections}

%\ccsdesc[500]{Computer systems organization~Embedded systems}
%\ccsdesc[300]{Computer systems organization~Redundancy}
%\ccsdesc{Computer systems organization~Robotics}
%\ccsdesc[100]{Networks~Network reliability}

%
% End generated code
%

%
%  Use this command to print the description
%
\printccsdesc

% We no longer use \terms command
%\terms{Theory}

\keywords{Collection; Data set; Online Social Networks}
\input{introduction}
\input{methods}
\input{statistics}
\input{conclusion}

%ACKNOWLEDGMENTS are optional
%\section{Acknowledgments}
%TODO: Do we put this section?
%This section is optional; it is a location for you
%to acknowledge grants, funding, editing assistance and
%what have you.  In the present case, for example, the
%authors would like to thank Gerald Murray of ACM for
%his help in codifying this \textit{Author's Guide}
%and the \textbf{.cls} and \textbf{.tex} files that it describes.

%
% The following two commands are all you need in the
% initial runs of your .tex file to
% produce the bibliography for the citations in your paper.
\bibliographystyle{abbrv}
\small
\bibliography{sigproc}
%\bibliography{sigproc.bib}
\normalsize
% sigproc.bib is the name of the Bibliography in this case
% You must have a proper ".bib" file
%  and remember to run:
% latex bibtex latex latex
% to resolve all references
%
% ACM needs 'a single self-contained file'!
%

%\subsection{References}
%Generated by bibtex from your ~.bib file.  Run %latex,
%then bibtex, then latex twice (to resolve %references)
%to create the ~.bbl file.  Insert that ~.bbl %file into
%the .tex source file and comment out
%the command \texttt{{\char'134}thebibliography}%.

% This next section command marks the start of
% Appendix 
\end{document}

%% file: introduction.tex
\section{Introduction}
\label{sec:introduction}

The field of computational social science and data-driven research is growing in importance~\cite{socialnetworkslazer2009}, and with this trend, there is a need for common academic benchmarking collections to facilitate a robust and reproducible research environment. In practice, however, datasets are often obtained via ad-hoc collection, or on the basis of proprietary data.

While originally Online Social Networks (OSNs) focused on allowing users to communicate, connect with others and share content, nowadays the term includes platforms which are primarily user-centric, allowing members to broadcast personal thoughts and content. \cite{browsingkumar2010} finds that OSNs are among the most frequently visited Web sites for a large population of users. In consequence, they can be used to study human behaviour at a large scale. Furthermore, because many OSNs are public by default and provide APIs to access their content, they have become good candidates for data collection to be used to study problems such as user/ topic modelling, user identification or information leakage.

In this paper, we introduce a collection of 850 users and their online footprint (part of their generated content and user profiles) spread across three social networks: Twitter, Instagram and Foursquare. It is our objective to provide a dataset on which privacy-sensitive methods and the defense against them can be tested. The construction of this dataset was initially carried out during a research project studying cross platform privacy loss arising from public information sharing on social networks~\cite{Hn2015}. In Table~\ref{table:datasets}, a comparison of our dataset with other existing corpora is given.

\begin{table*}
\centering
\caption{An overview of existing datasets containing cross-OSN user profiles.}
\label{table:datasets}
\begin{tabular}{|l|c|c|c|c|c|} \hline
Features& MNA~\cite{Mna} & MAH~\cite{MAH} & About.me~\cite{ASONAM2015} & NUS-MSS~\cite{Farseev2015} & Cross OSN\\ \hline
OSNs & \makecell{Twitter, \\ Foursquare} & \makecell{Twitter, \\ BlogCatalog} & \makecell{ Flickr, Google+, Instagram,\\ Tumblr, Twitter, Youtube} & \makecell{Instagram, Twitter \\ Foursquare} &\makecell{Instagram, Twitter \\ Foursquare}\\ \hline
\# Users & 500 & 2710 & 15,595 & 20,483 & 850\\ \hline
Content type & \makecell{User IDs, posts,\\ friends graph} & \makecell{User IDs, \\ friends graph} & \makecell{User IDs,\\ post IDs} & \makecell{Anonymised \\ timeline data} & \makecell{User IDs, \\ post IDs}\\\hline
Availability & Unknown & Unknown & Available & Available & Available\\
\hline\end{tabular}
\end{table*}

Our data collection method relies on linking users across three popular OSN platforms. Twitter is a microblogging platform whose main content comes in \emph{tweets}, posts limited to 140 characters which can contain text, video or images, links to external Web sites, references to other users and \emph{hashtags} (terms starting with the \# symbol used to mark keywords or topics in a tweet). Instagram is a photo sharing platform. Its main content are photos or videos along with optional textual descriptors. Foursquare is a location service platform concentrating on the notion of \emph{check-ins}.
%\footnote{Check-ins correspond to real-world venues that the user has visited. In addition to the venue name, more information such as location and venue types are available.}.

%In this paper we present a unique data set which comprises of triplets of profiles across the aforementioned OSNs of over 1000 users.

The paper makes the following novel contributions (1) We describe the methodology and release the code to construct a user-centric cross-OSN dataset. (2) We release a dataset of 850 profile triples across the aforementioned OSNs.

The remainder of this paper is structured as follows: in Section~\ref{sec:method}, we present the methodology used to create the collection. Section~\ref{sec:statistics} presents key statistics and qualitative aspects of the dataset. Section~\ref{sec:discussion} discusses the task of cross-platform user identification as an example use case before sketching a range of further conceivable use cases and tasks.

%% file: methods.tex
\section{Method}
\label{sec:method}
We use Twitter, Instagram and Foursquare mainly due to the ease of crawling and publicly accessible APIs, but also because the content generated by the users in these distinct social networks is diverse and representative of a comprehensive range of OSN use cases. In~\cite{twittercontent2012}, the authors find that on Twitter, the top types of content users share are: personal information, random thoughts, opinions/complaints and facts (\textit{e.g.} news). While on Instagram, the majority of the posted pictures can be put in 1 of 8 categories, with the leading categories being \emph{selfies} and \emph{friends}~\cite{instagramcontent2014}. On Foursquare, users share their location in terms of venues, which carries not only information in the form of raw geographic coordinates, but often also the venue's name and function. 

\begin{figure*}
\centering
\includegraphics[height=3.2in]{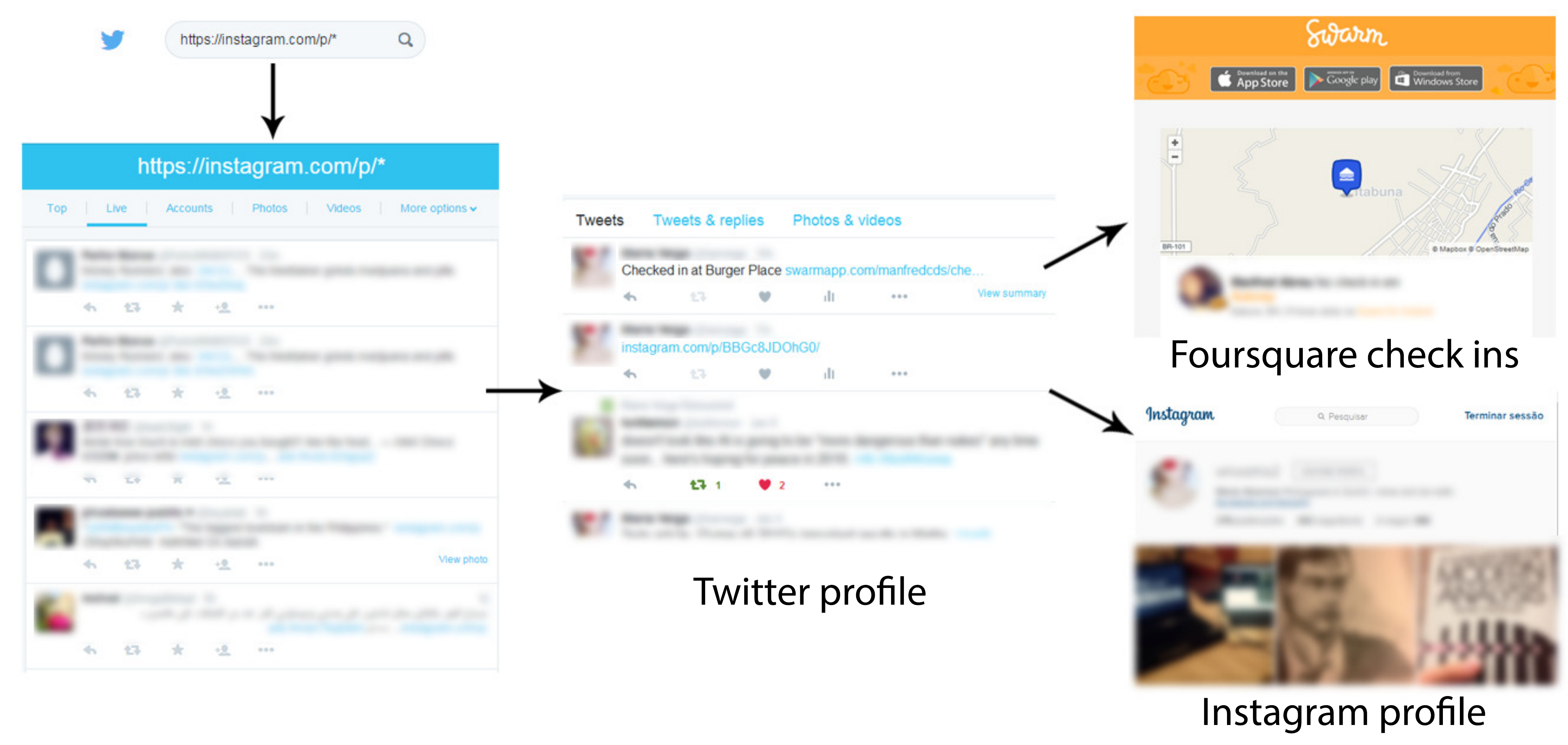}
\caption{Dataset collection. We use Twitter's search API to find posts with the pattern of a Foursquare check-in or an Instagram post. If a profile contains both, we crawl the Foursquare check-ins and Instagram profile through their respective APIs.}
\label{fig:dataset}
\end{figure*}

In order to obtain profiles from different OSNs belonging to the same person, we first use the Twitter Search API to search for specific post patterns (\textit{e.g.}, \url{https://instagr.am/p/*}), 
in order to identify users who cross-post content from other platforms on Twitter. A graphical depiction of the collection process is shown in Figure~\ref{fig:dataset}.

The data was collected in January and February 2016. While we were able to find over 5000 profile triples, only approximately 20\% of these triples fulfill our criteria of actively using the three selected OSNs, posting predominantly in English and sharing content publicly. After enforcing these requirements, our dataset contains a total of 850 distinct user profiles.

\subsection{English predominance}
We focused on profiles with predominantly English content as the initial study for which the data was collected was a natural language processing task that would have suffered from excessive amounts of cross-language content.

In order to guarantee that the majority of the content is posted in English, we use the method described in Algorithm~\ref{alg:english}. We set the ratio to be $0.1$ and $K=100$ for our collection. The purpose is not to exclude users that occasionally post in a non-English language, but to make sure the data set does not contain too many strictly non English-speaking users.

\begin{algorithm}
\SetAlgoLined
\KwData{twitter timeline}
\KwResult{true or false}
initialization\;
\For{K posts}{
    count(english posts)\;}
    
\eIf{ \mbox{count.total}/K < ratio }{
    disregard user\;
   }{
  crawl timeline\;
  }
 \caption{English check}
 \label{alg:english}
\end{algorithm}

\subsection{Spam detection}
\label{sec:spammer}
To increase the confidence that users are authentic personal accounts instead of spammers that merely redistribute content from other users, we have a simple heuristic described in Algorithm~\ref{alg:identity}. We take the timeline of the user from the respective OSN and we check whether the majority of the posts come from the same user name. In our collection, the threshold is set to 30\%. Suspected spammer triples are flagged but, for completeness, remain in the dataset. 

\begin{algorithm}
\SetAlgoLined
\label{alg:identity}
 \KwData{Instagram or Foursquare timeline}
 \KwResult{spammerFlag}
 initialization\;
 \For{posts}{
  count(author of post)\;}
  
  \eIf{ counter.max/total posts < threshold }{
   spammerFlag=1\;
   }
  {spammerFlag=0\;
  }
 \caption{Identity check}
\end{algorithm}

%% file: statistics.tex
\section{Dataset}
\label{sec:statistics}
%In this section we present information about the dataset.

%\subsection{Statistics}
Our 850 authentic English-speaking users produced approximately 2.5M tweets, 340k check-ins and 42k Instagram posts\footnote{\url{http://cake.da.inf.ethz.ch/OSN-sigir2016/}}. Four users were flagged as spammers under the rule described in Section~\ref{sec:spammer}.

\begin{figure}
\centering
\includegraphics[height=2in]{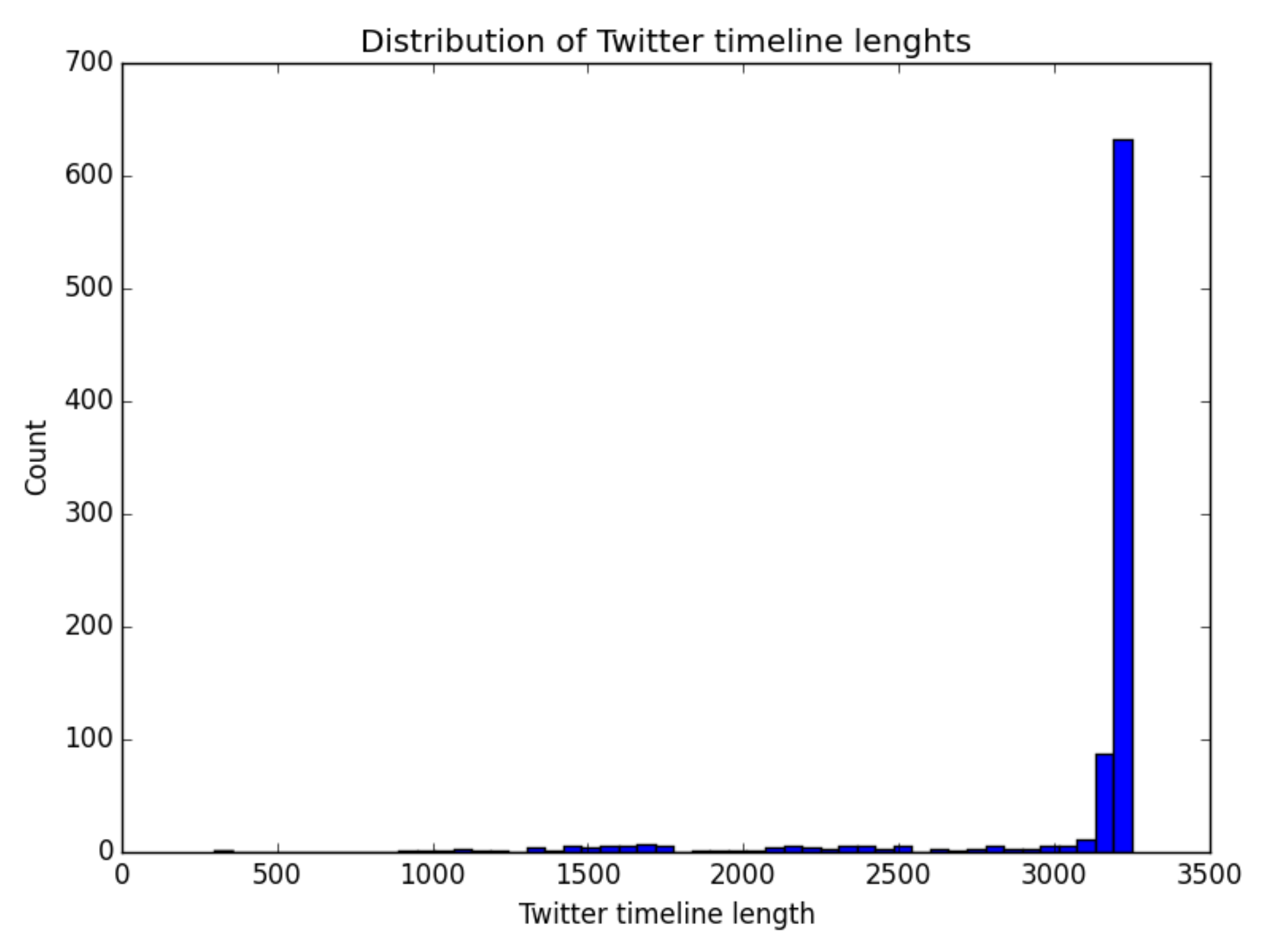}
  \caption{Histogram of tweets per user}\label{fig:twitterdist}
\end{figure}

\begin{figure}
\centering
\includegraphics[height=2in]{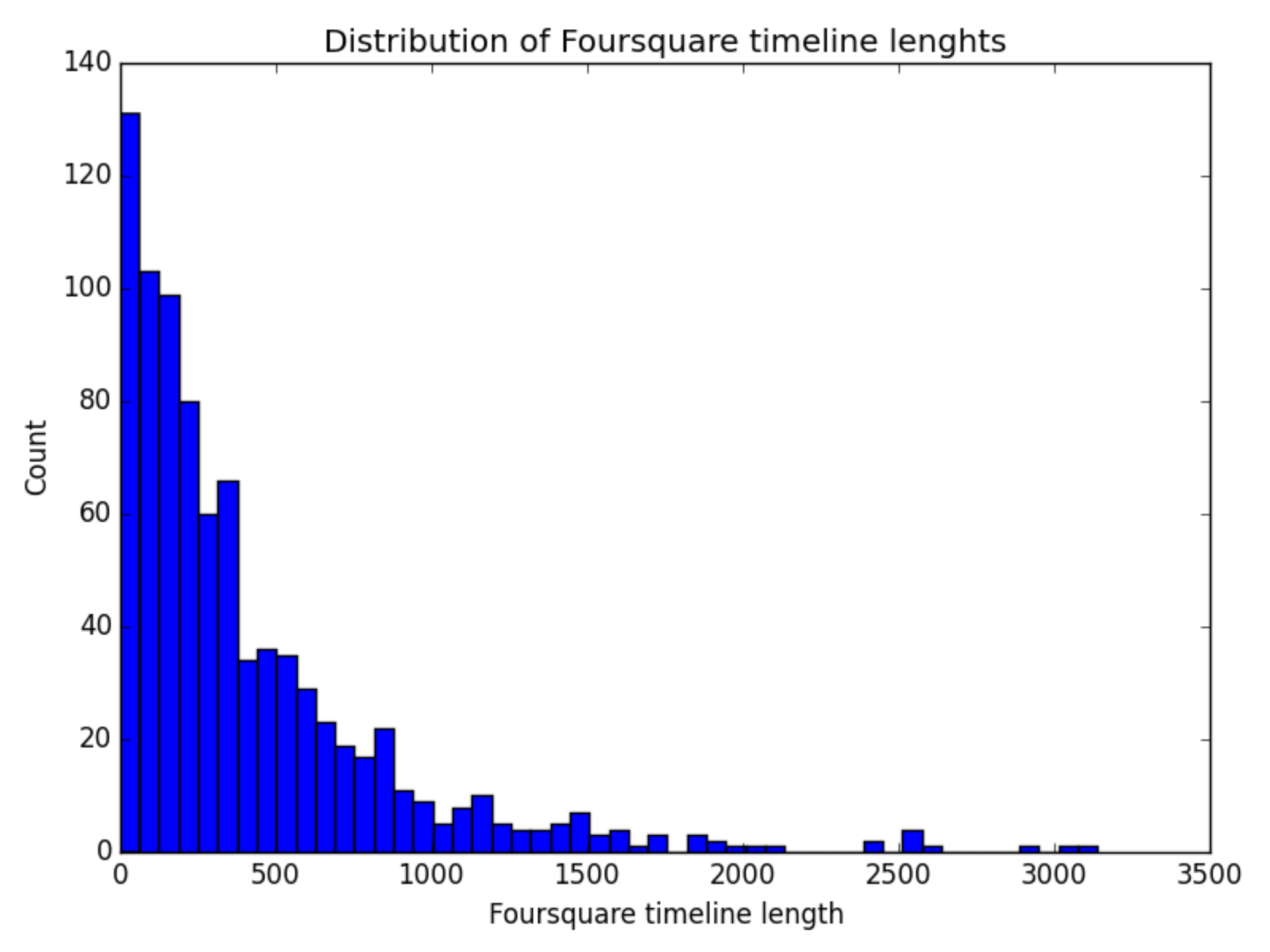}
  \caption{Histogram of Foursquare posts per user}\label{fig:4sqdist}
\end{figure}

\begin{figure}
\centering
\includegraphics[height=2in]{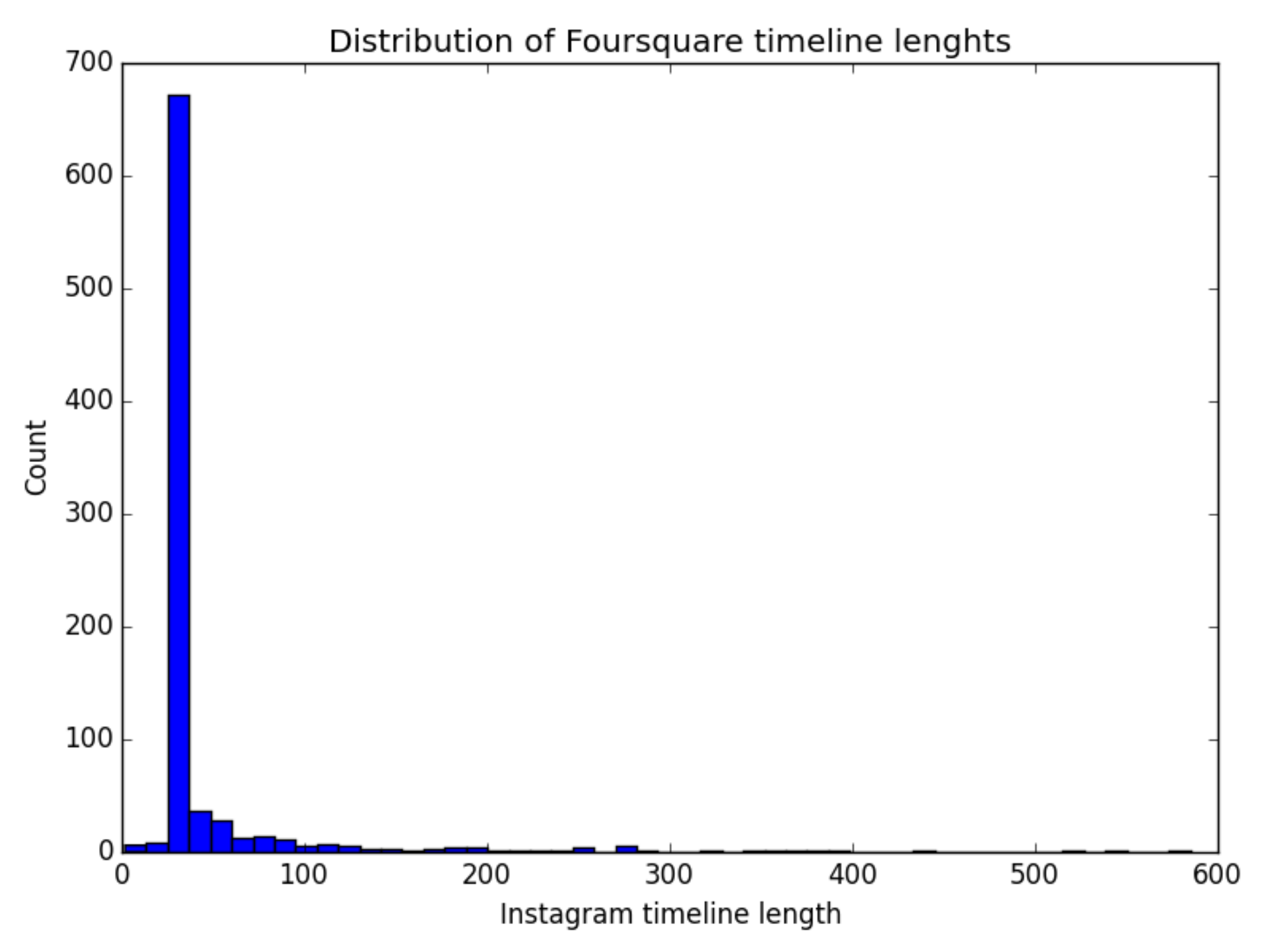}
  \caption{Histogram of Instagram posts per user}\label{fig:instdist}
\end{figure}

Figure~\ref{fig:twitterdist} shows the distribution of tweets per user. Figures \ref{fig:4sqdist} and \ref{fig:instdist} show the distribution of number of check-ins and Instagram posts per user, respectively.

The Twitter API restricts access to at most 3200 tweets per profile (including re-tweets)~\cite{twitterAPI}. Because we exclude direct re-tweets from our data set, the majority of the Twitter profiles we collect contain between 3000 and 3200 tweets. For each Instagram profile, we recover the most recent posts through the Instagram API and complement them with existing content which has been posted on Twitter. For each Foursquare profile, we recover those check-ins that were cross-posted on the retrieved Twitter timeline.

\begin{figure}
\centering
\includegraphics[height=2in]{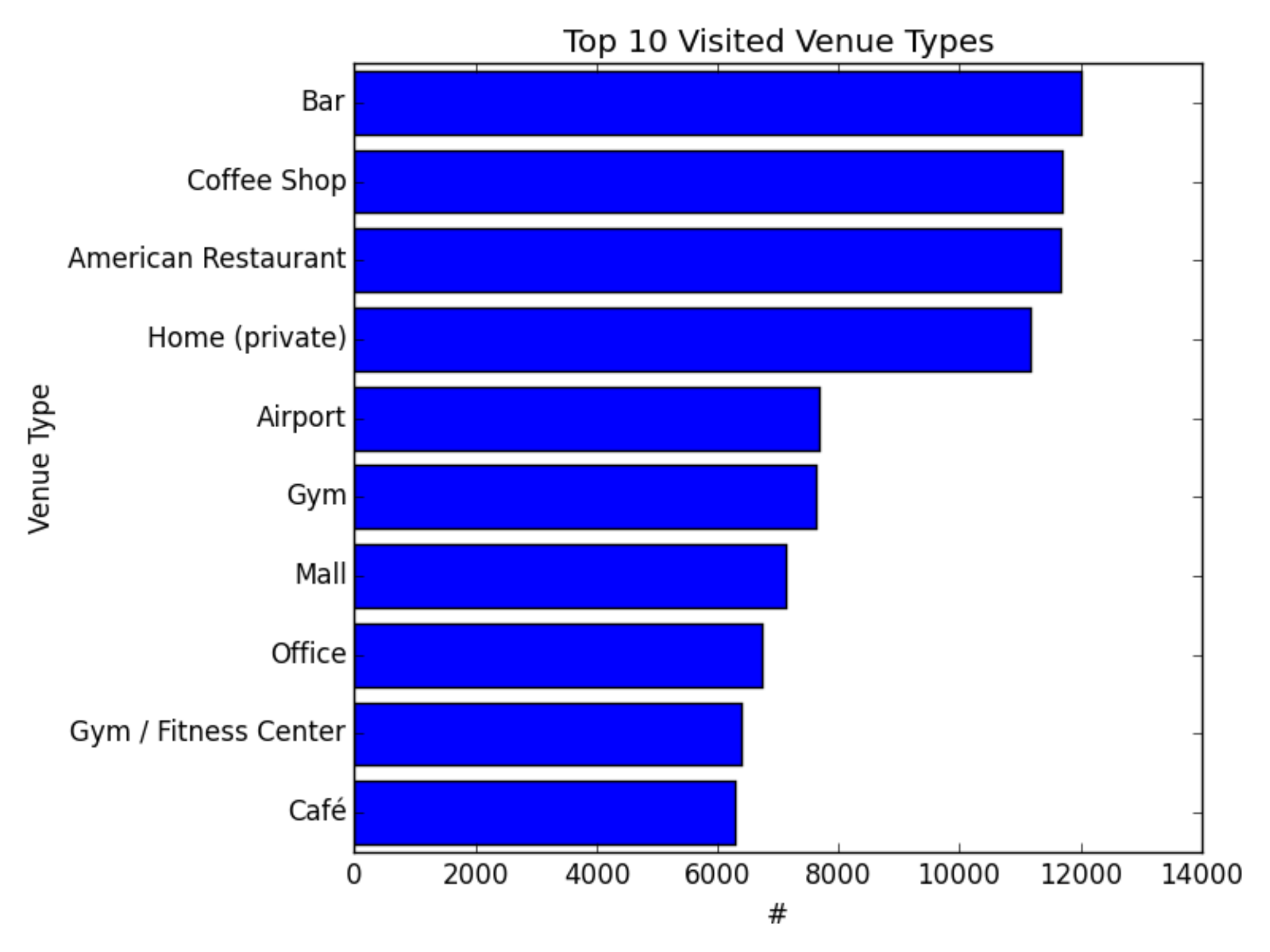}
  \caption{Top 10 visited venue types}
  \label{fig:mostvisited}
\end{figure}

In this dataset, we observe check-ins from 111 countries, spawning 669 venue types. The most visited venue types are shown in Figure~\ref{fig:mostvisited}. %The content in these timelines spans from 2007 to 2016.%, as shown in Figure~\ref{fig:timespan}.

%Total number of check-ins: 383469
%Total number of cities: 6941
%Total number of countries: 106
%Total number of venue types: 656

%\begin{figure}
%\centering
%\includegraphics[height=2in]{timespan}
%  \caption{Time span from collected content}
%  \label{fig:timespan}
%\end{figure}

\subsection{Properties}
 For each user we store Twitter, Instagram and Foursquare IDs along with the tweet and post IDs of all shared content. The Foursquare data is represented as a list of tuples $(venue\_id,tweet\_id)$, where \emph{tweet\_id} is the tweet announcing the check-in. An example of the collected user data is shown in Table~\ref{table:userexample}. For copyright reasons, we do not distribute the content but instead provide the scripts to crawl the data~\cite{github}. Using these scripts, the following can be retrieved:
\begin{itemize}
    \item From Twitter: using the user ID, a profile object that contains information such as the name, location, date of creation of account, and with the post ID, a the tweet object, containing the tweet and metadata~\cite{twitterAPI}.
    \item From Instagram: using the user ID, a profile object containing information such as the name, location, date of creation of account, and with the post shortcode from Instagram, an Instagram object containing the link with the image and related metadata~\cite{instagramAPI}.
    \item From Foursquare: using the venue ID, the venue object can be retrieved, containing information such as venue type, venue name and location~\cite{4sqAPI}.
\end{itemize}

\begin{table}
\centering
\caption{Example of provided and retrievable data.}
\label{table:userexample}
\begin{tabular}{|l|l|l|} \hline
{\bf Feature}&{\bf Description}& {\bf Retrieves}\\ \hline
\multicolumn{3}{l}{\textsc{Twitter}\parbox{0pt}{\rule{0pt}{1ex}}}\\[2pt] \hline
%Twitter & & \\ \hline
ID & Integer & Twitter profile\\ \hline
Timeline & List of tweet IDs & Tweet object\\ \hline
\multicolumn{3}{l}{\textsc{Instagram}\parbox{0pt}{\rule{0pt}{1ex}}}\\[2pt] \hline
ID & Integer & Instagram profile\\ \hline
Timeline & List of post shortcodes & Instagram object\\ \hline
 \multicolumn{3}{l}{\textsc{Foursquare}\parbox{0pt}{\rule{0pt}{1ex}}}\\[2pt] \hline
ID & User unique identifier & -- \\ \hline
Timeline & \makecell{List of pairs \\ (tweet\_id,venue\_id)} & Venue object\\
\hline\end{tabular}
\end{table}

%% file: conclusion.tex
\section{Discussion}
\label{sec:discussion}
In this section we present a use case for this dataset in a cross-OSN user identification task. Then, we elaborate on other tasks that can find this collection useful.

\subsection{User Identification}
The task of user identification is concerned with matching profiles from different domains belonging to the same natural person~\cite{Zafarani2013}. In this scenario, we use Instagram and Twitter as our data sources. A simple, yet powerful baseline algorithm is used, which compares the similarity of the nicknames on the respective platforms.

By assigning minimal Levenshtein distance between user names and choosing the one with the lowest edit distance, we attain a matching accuracy of 70.1\%. If the user names are preprocessed by converting them to lowercase (as in both these social networks the letter case does not make a difference), we attain an accuracy of 72.8\%. One can think of many more advanced schemes tracking common topics or writing styles across social networks. 

\subsection{Content generation and spreading}
In~\cite{pinsandtweets}, the authors study how users behave across Pinterest and Twitter. Other papers study Twitter or Instagram alone~\cite{twittercontent2012,instagramcontent2014}. It would be interesting to study the behaviour of users in these OSNs while having access to their multiple profiles. This could help answer whether some topics are OSN-specific or whether the activity in one profile can indicate future activity in another profile.

\newpage
\subsection{Cross-OSN inference}
Following the idea that an activity on one platform can indicate future activity on another one, an interesting study would be to see whether it is possible to infer information from one OSN regarding another one. For example, whether it is possible to predict topics, interests or intentions. An example of this type of work can be found in~\cite{Hn2015}, where the authors use Twitter timelines to infer venue type visits.

\subsection{Aggregated OSN topic modelling}
Because OSNs can carry different information depending on their functionality, the access to several profiles from the same user across different platforms might give an advantage when modelling the user. These user models can be used for targeted advertisement or recommender systems. An interesting task would be to compare whether the inclusion of different profiles can benefit the models or not.

%\subsection{Further discussion}

\section{Conclusion}
In this paper, we described the collection, structure and properties of a benchmarking corpus of cross-platform OSN user profiles useful for a wide range of privacy-related research questions. It encompasses hundreds of user triples, millions of tweets and thousands of Instagram and Foursquare posts. We ensure a focus on English-speaking users and perform spammer identification to reduce noise in the dataset. To the best of our knowledge, this dataset is the first of its kind both in nature as well as scale. 
%. In order to reduce the inherent amount of noise in the dataset, we ensure a focus on English-speaking users and perform spammer identification. %To the best of our knowledge, this dataset is the first of its kind both in nature as well as scale.  